# Simple Demonstration of Different Types of Coupling in Multiphysics Numerical Problems

Alexandre Lavrov

*Norwegian University of Science and Technology, Trondheim, Norway*

Email: alexandre.lavrov@ntnu.no

Numerical modelling of coupled multiphysics phenomena is becoming an increasingly important subject in applied mathematics. The main challenge in teaching this subject is the complexity of both the mathematical models and their numerical implementation. In this note, a simple demonstrator is proposed that enables demonstration of and some hands-on experience with three types of coupling commonly used in computational science: one-way coupling, explicit sequential coupling, and full coupling. It makes use of a familiar nonlinear heat equation in 1D, with the solution being a propagating heat wave.

## 1. Introduction

Numerical modelling in today's engineering sciences often involves multiphysics problems where several physical processes are coupled together (Zhang & Cen, 2015). Each of these processes is described by its own set of equations, and the solution of either of them enters the other(s) as input data. Examples of coupled processes are found in different areas of engineering. For instance, fluid-structure interaction takes place wherever there is a mobile and/or deformable solid component immersed in a fluid (Paidoussis, 2014). An example would be interaction between a ship and a wave. Here, the fluid flow affects the deformation and displacement of the solid body. This, in turn, changes the boundary conditions for the fluid flow and thereby affects the flow. Many examples of coupled systems are found in geosciences, e.g. in geothermics (Wang et al., 2022). In geothermal systems, pressure and temperature changes within the reservoir affect the stress state. This, in turn, affects the fracture properties, which alters the reservoir permeability. This alters the fluid flow and subsequently the temperature distribution, which again affects the stress state. In oil production, extracting oil from the reservoir affects the stress state and induces rock deformation (Fjær et al., 2008). This deformation alters the rock permeability, which, in turn, affects the fluid flow. In tunnel engineering, grouting of weak and/or permeable rock masses is often used in order to enable construction of a tunnel. Grouting means that cement is injected into natural fractures in order to make them less permeable. During injection, fluid pressure inside the fractures rises, thereby increasing the fracture aperture (opening). Increased aperture affects the fluid flow through the fracture system (Saeidi et al., 2013). Coupling between fluid flow in fractures and mechanical stresses acting in the rock is also a key feature in hydraulic fracturing, a technique used to stimulate oil wells and to perform stress measurements in rocks (Lavrov, 2017). Finally, our last example will be from underground $CO_2$ storage which is currently considered as a viable solution to the greenhouse gas problem. Injection of $CO_2$ into a geological formation for long-term storage affects the stress state (Vilarrasa et al., 2010). It also may induce chemical interactions between $CO_2$ and the formation fluids (and the host rock, too). This, in turn, will affect the formation permeability and thus injectivity of the injection wells.

As engineering projects become more complex, the prevalence and importance of coupled phenomena to be modelled by tomorrow's engineers is going to increase. Therefore, it is crucial that essentials of coupled modelling be explained in sufficient detail as part of university courses on numerical modelling.

At mathematical modelling level, coupled models encountered in engineering sciences are represented as systems of partial differential equations, e.g. fluid flow equations, solid mechanics equation and heat conduction equation. At implementation level, these equations are solved either by using a single, advanced software package or by coupling together several simpler software packages each of which takes care of its own physics. This complexity of implementation easily obnubilates the essential principles behind coupled modelling. Therefore, there is a need for a demonstrator that could highlight essential principles of couped modelling, at the same time keeping the mathematical and implementation details relatively simple and transparent. The objective of this note is to develop such a demonstrator. As we shall see, the demonstrator allows us to illustrate, in particular, the distinction between three types of coupling: one-way coupling, full coupling, and sequential coupling, all introduced in Section 2.

## 2. Types of coupling

As mentioned above, a coupled problem is essentially a set of partial differential equations where solution of one or several of them enters the other(s). Several strategies for numerical solution of such systems have been developed over the last decades. They are summarized in Figure 1.

A multiphysics system may involve any number of processes such as fluid flow, heat transfer, solid deformations, chemical reactions etc. In this Section, we will explain the difference between the coupling types by means of a simple system where only two physical processes are in play and, thus, the system is described by two sets of equations. We call these processes A and B. For instance, in fluid-structure interaction A could be fluid flow, B could be deformation and motion of the solid components. Process A affects parameters and sources in process B. Process B affects parameters and sources in process A. Thus, in order to establish a coupled model, two sets if equations need to be solved simultaneously: the equations governing process A and those governing process B.

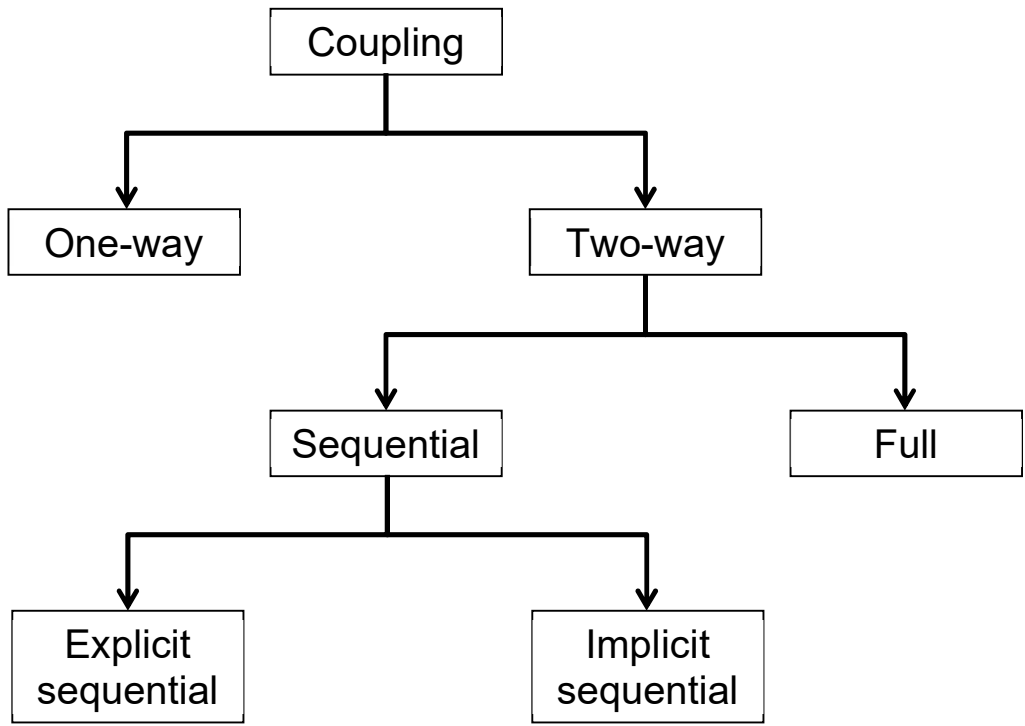

Figure 1　　Classification of different coupling methods.

The 'cheapest' strategy, from computational point of view, is known as one-way coupling. In this scenario, transfer of information between the coupled sets of equations is one-way only. For instance, the output from solving the equations describing process A is to be used in the equations describing process B but the results of solving the equations describing process B are never passed back to the process A. Accuracy of one-way coupling is poor.

Exchanging information between processes A and B in both directions is known as two-way coupling and can be achieved in a number of ways (Figure 1). One option is to consider the two sets of equations as one set. The equations can then be discretized using available approximation techniques, e.g. finite-difference or finite-element methods. The resulting full system of linear or nonlinear algebraic equations can then be solved simultaneously at subsequent timesteps. This approach is known as full coupling. This type of coupling is expected to produce the most accurate results. In practice, solving a large system of, in general, nonlinear partial differential equations may result in unsurmountable problems. An additional drawback is that it results in a code design that is quite rigid.

Between these two extremes, i.e. inaccurate but simple one-way coupling and accurate but complicated full coupling, lies the third option, the sequential coupling (Figure 2). According to this strategy, the two sets of equations are solved separately, and information is transferred between them at each timestep (Tran et al., 2004). Separate modules are thus required to solve equations describing process A and equations describing process B. This modularity is a strength of sequential coupling. Another strength is that the equations to be solved in each of the modules are considerably simpler than the set of equations assembled in fully coupled models. The two sequentially coupled modules can be implemented as parts of a single code, or two (or more) codes can be coupled together, even codes delivered by different software vendors.

Sequential coupling can be achieved in two different ways. One option is to exchange information only once per timestep. For instance, one computes process A, sends the results to process B, computes process B, sends the results to process A that then proceeds to the next timestep. This option is known as explicit sequential coupling (Rutqvist et al., 2002). Another option is to run the two simulators (one for process A and one for process B) at each timestep in loop, until convergence is reached. This is known as implicit sequential coupling (Rutqvist et al., 2002). There is evidence that, given a sufficiently small timestep, implicit sequential coupling produces the same results as full coupling. Explicit sequential coupling is less computationally expensive than implicit sequential coupling because, in the former, both modules, A and B, are run only once at each timestep. When implicit coupling is used, both modules are run in loop until convergence is reached at each timestep. Furthermore, there are other variations of sequential coupling where information exchange between A and B modules is performed at selected timesteps only (Dean et al., 2006), thereby rendering this coupling method even more like the one-way coupling.

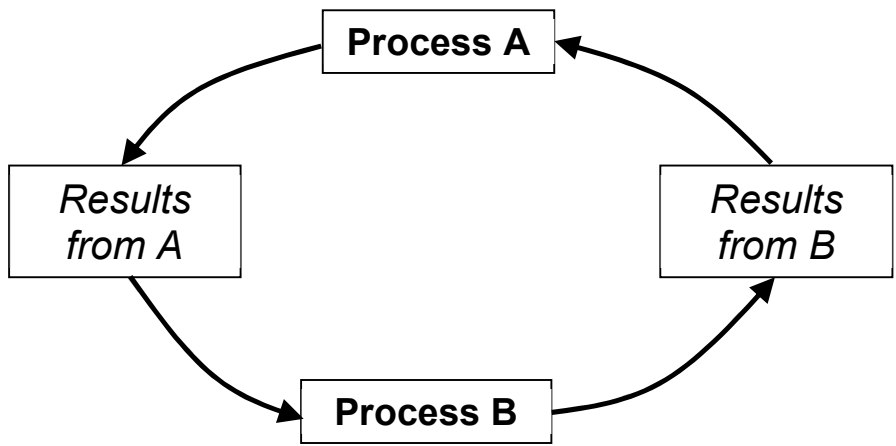

Figure 2    Sequential coupling in a multiphysics simulation. Blocks in italics show parameters passed between the modules that solve two sets of differential equations describing physical processes A and B (shown in bold).

It should be noted that, unfortunately, there is no standard terminology for different types of coupling. For instance, one-way coupling is sometimes termed 'explicit'. Full coupling is sometimes termed 'implicit', 'strong' or 'direct'. Sequential coupling is sometimes termed 'iterative'. Explicit sequential coupling is sometimes termed 'loose'. Implicit sequential coupling is sometimes termed 'strong'. References (Dean et al., 2006; Kim et al., 2018; Minkoff et al., 2003; Zhang & Cen, 2015) present a variety of terms for the coupling methods. The names used in Figure 1 have the appeal of being quite descriptive but they are by no means universally accepted or standard.

**3. Demonstrator for three types of coupling**

A relatively simple problem from numerical heat transfer can be used to illustrate differences between different types of coupling, and between the simulation results they produce. Consider one-dimensional heat transfer with the thermal diffusivity, $D$, being a function of temperature:

$$\frac{\partial T}{\partial t} = \nabla(D\nabla T), \; x \in [0, \; 1]; \; T(0,t) = T_l; \; T(1,t) = T_r; \; T(x,0) = T_r \qquad (1)$$

where $T_l$ and $T_r$ are temperature values at the left and right boundary, respectively. In particular,

$$D(T) = \gamma T^a \qquad (2)$$

with $\gamma$ being a constant and $a$ being equal to 3 describes radiation heat transfer (Hammer & Rosen, 2003; Marshak, 1958).

Eqs (1)-(2) are solved numerically by means of the finite-difference method using three types of coupling: full coupling, explicit sequential coupling, and one-way coupling. Full coupling is obtained by substituting $D$ from Eq. (2) into Eq. (1), and discretizing the resulting differential equation using e.g. the BTCS (backward in time, central in space) scheme (Thomas, 1995). The only subtlety here is the treatment of internodal diffusivity. In our demonstrator, we use arithmetic averaging of nodal diffusivities in order to approximate internodal diffusivity. This was shown in (Kadioglu et al., 2008) to yield superior results on heat wave propagation problems like the one at hand. The resulting finite-difference scheme with full-coupling is thereby given by:

$$f = T_k^{(n+1)} - T_k^{(n)} - \frac{\gamma \Delta t}{2 \Delta x^2} \left\{ \left[T_{k+1}^{(n+1)}\right]^{a+1} - \left[T_{k+1}^{(n+1)}\right]^a T_k^{(n+1)} + \left[T_k^{(n+1)}\right]^a T_{k+1}^{(n+1)} - 2\left[T_k^{(n+1)}\right]^{a+1} + \left[T_k^{(n+1)}\right]^a T_{k-1}^{(n+1)} - \left[T_{k-1}^{(n+1)}\right]^a T_k^{(n+1)} + \left[T_{k-1}^{(n+1)}\right]^{a+1} \right\} = 0$$

(3)

with the Jacobian matrix given by

$$J_{k,k-1} = \partial f / \partial T_{k-1}^{(n+1)} = -\frac{\gamma \Delta t}{2\Delta x^2}\left\{\left[T_k^{(n+1)}\right]^a - a\left[T_{k-1}^{(n+1)}\right]^{a-1} T_k^{(n+1)} + (a+1)\left[T_{k-1}^{(n+1)}\right]^a\right\}$$

$$J_{k,k} = \partial f / \partial T_k^{(n+1)} = 1 - \frac{\gamma \Delta t}{2\Delta x^2}\left\{-\left[T_{k+1}^{(n+1)}\right]^a + a\left[T_k^{(n+1)}\right]^{a-1} T_{k+1}^{(n+1)} - 2(a+1)\left[T_k^{(n+1)}\right]^a + a\left[T_k^{(n+1)}\right]^{a-1} T_{k-1}^{(n+1)} - \left[T_{k-1}^{(n+1)}\right]^a\right\}$$

$$J_{k,k+1} = \partial f / \partial T_{k+1}^{(n+1)} = -\frac{\gamma \Delta t}{2\Delta x^2}\left\{\left[T_k^{(n+1)}\right]^a - a\left[T_{k+1}^{(n+1)}\right]^{a-1} T_k^{(n+1)} + (a+1)\left[T_{k+1}^{(n+1)}\right]^a\right\}$$

. (4)

Here, the parenthesized superscripts, e.g. ($n$), denote the timestep. The subscripts, e.g. $k$, denote the node number.

The system of Eqs. (3) is solved by Newton-Raphson method at each timestep. Newton-Raphson iterations are run until 1-, 2-, and ∞-norms of the residual became smaller than $10^{-6}$ and 1-, 2-, and ∞-norms of relative changes of the solution between iterations became smaller than $10^{-6}$.

Explicit sequential coupling is achieved by using temperature values from the previous time step in the diffusivity when discretizing Eq. (1). Treating, again, the internodal diffusivity as arithmetic average of its nodal values, the following scheme is obtained:

$$-\frac{\gamma \Delta t}{2\Delta x^2}\left\{\left[T_{k-1}^{(n)}\right]^a + \left[T_k^{(n)}\right]^a\right\} T_{k-1}^{(n+1)} + \left\{1 + \frac{\gamma \Delta t}{2\Delta x^2}\left\{\left[T_{k+1}^{(n)}\right]^a + 2\left[T_k^{(n)}\right]^a + \left[T_{k-1}^{(n)}\right]^a\right\}\right\} T_k^{(n+1)} - \frac{\gamma \Delta t}{2\Delta x^2}\left\{\left[T_{k+1}^{(n)}\right]^a + \left[T_k^{(n)}\right]^a\right\} T_{k+1}^{(n+1)} = T_k^{(n)}$$

(5)

The computational task is now considerably lighter and the execution is faster because the scheme is a system of linear equations.

Finally, the third scheme is based on one-way coupling. Here, the diffusivity in Eq. (1) is never updated for the new values of temperature but the diffusivity is computed separately in Eq. (2) using the updated values of temperature. Eq. (1) is thus solved using the textbook BTCS scheme for linear heat equation:

$$-\frac{D\Delta t}{\Delta x^2} T_{k-1}^{(n+1)} + \left(1 + \frac{2D\Delta t}{\Delta x^2}\right) T_k^{(n+1)} - \frac{D\Delta t}{\Delta x^2} T_{k+1}^{(n+1)} = T_k^{(n)} \quad . \tag{6}$$

The three schemes outlined above are implemented in Python using Numpy libraries. The implementation takes less than 300 code lines.

Figures 3 and 4 illustrate a comparative performance of the three schemes presenting, respectively, the computed temperature profiles and diffusivity profiles at target time 0.5. Three panels are given in each Figure; these present results obtained with time step 0.001, 0.005, and 0.01. The following input data were used in the simulations: number of nodes (including the boundary nodes): 200; $T_l = 0.1$; $T_R = 2.0$; $a = 3.0$; exit tolerance in the BiCGSTAB solver in NumPy when using one-way and explicit sequential coupling: $10^{-7}$.

The fully coupled scheme produces, at all timestep values, the expected classical result, i.e. a thermal shock wave propagating in the positive direction of $x$, the so-called 'Marshak wave' (Kadioglu et al., 2008).

Figure 3 demonstrates that the results obtained with explicit sequential coupling are relatively close to the expected solution at the smallest time step (Figure 3a) but deteriorate rapidly as the timestep increases. The results obtained with explicit sequential coupling become inferior at timestep 0.01 whereas the fully coupled scheme produces virtually the same results at all three timesteps. This behaviour is a common feature of explicitly coupled models.

Results produced with one-way coupling are inferior to the other two schemes at all three timestep values.

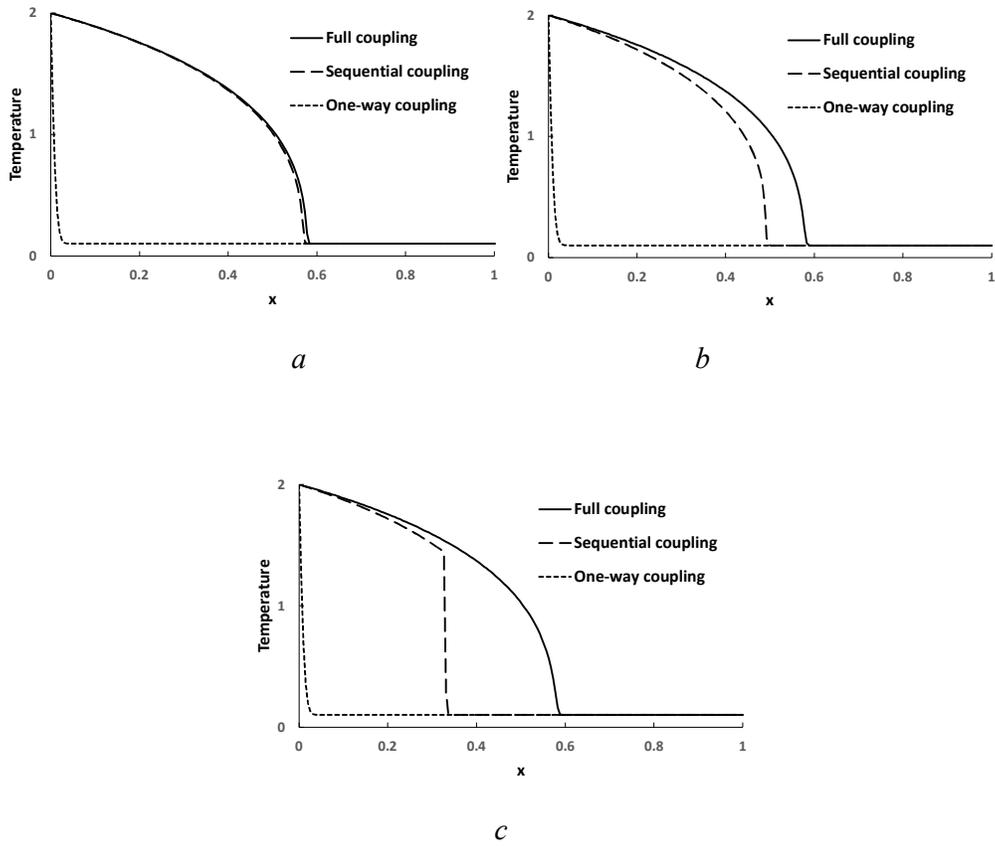

Figure 3    Temperature profiles at $t = 0.5$ obtained with the three coupling schemes using three timestep values: (a) 0.001; (b) 0.005; (c) 0.01.

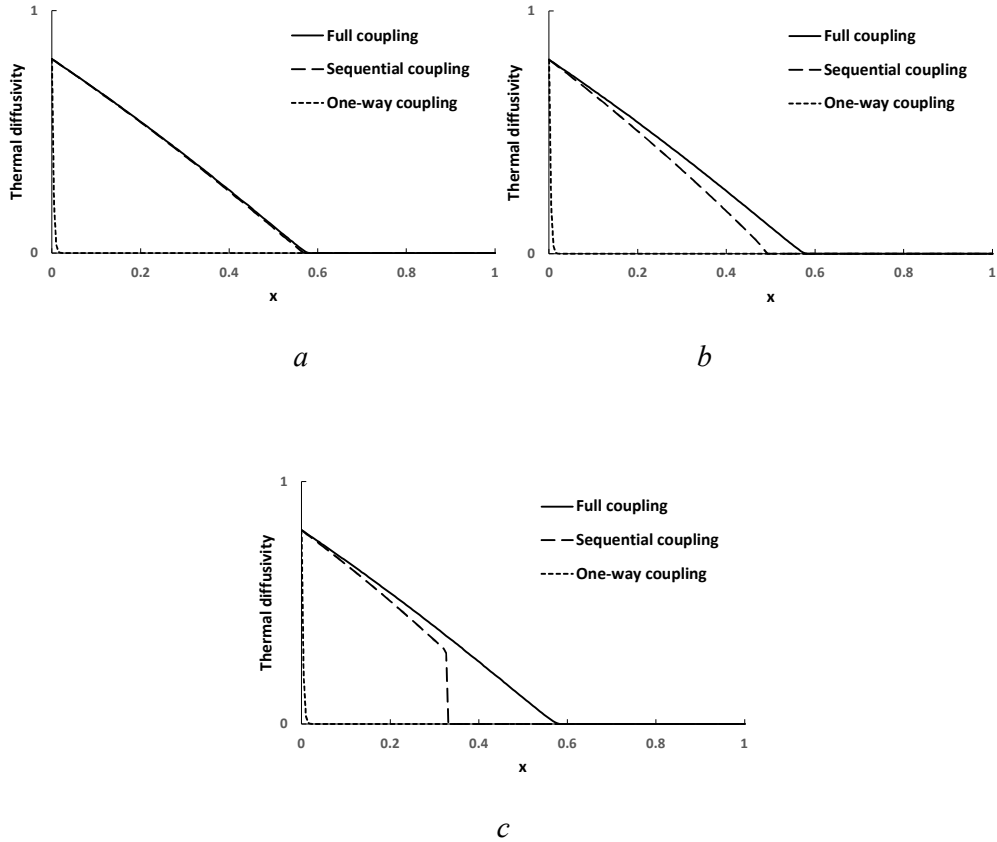

Figure 4    Thermal diffusivity profiles at $t = 0.5$ obtained with the three coupling schemes using three timestep values: (a) 0.001; (b) 0.005; (c) 0.01.

## 4. Conclusions

The proposed demonstrator is easy to implement in a high-level programming language such as Python. The demonstrator enables us to illustrate important features of a rather advanced topic in computational science, namely different types of coupling in multiphysics problems. In particular, the differences between different types of coupling, particularly full and sequential, become transparent because the equations to be coupled are the familiar 1D nonlinear heat equation and an algebraic equation used for diffusivity update. The demonstrator highlights typical features of the different types of coupling, in particular the superior accuracy and the need for iterations in the fully coupled model. It also shows that the accuracy of a sequentially coupled scheme deteriorates rapidly with increasing timestep ultimately producing an inferior solution, while the fully coupled scheme still performs adequately.

**Disclosure statement**

There are no competing interests to declare.

**Funding**

The support of the EU H2020 Marie Sklodowska-Curie grant no. 101008140 EffectFact is gratefully acknowledged.

Rutqvist, J., Wu, Y. S., Tsang, C. F., & Bodvarsson, G. (2002). A modeling approach for analysis of coupled multiphase fluid flow, heat transfer, and deformation in fractured porous rock. *International Journal of Rock Mechanics and Mining Sciences*, *39*(4), 429-442. https://doi.org/https://doi.org/10.1016/S1365-1609(02)00022-9

Saeidi, O., Stille, H., & Torabi, S. R. (2013). Numerical and analytical analyses of the effects of different joint and grout properties on the rock mass groutability. *Tunnelling and Underground Space Technology*, *38*, 11-25. https://doi.org/https://doi.org/10.1016/j.tust.2013.05.005

Thomas, J. W. (1995). *Numerical Solution of Partial Differential Equations: Finite Difference Methods*. Springer.

Tran, D., Nghiem, L., & Settari, A. (2004). New iterative coupling between a reservoir simulator and a geomechanics module. *SPE Journal*, *9*(03), 362-369. https://doi.org/10.2118/88989-pa

Vilarrasa, V., Bolster, D., Olivella, S., & Carrera, J. (2010). Coupled hydromechanical modeling of CO2 sequestration in deep saline aquifers. *International Journal of Greenhouse Gas Control*, *4*(6), 910-919. https://doi.org/https://doi.org/10.1016/j.ijggc.2010.06.006

Wang, H., Liu, H., Chen, D., Wu, H., & Jin, X. (2022). Thermal response of the fractured hot dry rocks with thermal-hydro-mechanical coupling effects. *Geothermics*, *104*, 102464. https://doi.org/https://doi.org/10.1016/j.geothermics.2022.102464

Zhang, Q., & Cen, S. (2015). *Multiphysics Modeling (Tsinghua University Press Computational Mechanics Series)*. Academic Press.